# Improved Method for Dealing with Discontinuities in Power System Transient Simulation Based on Frequency Response Optimized Integrators Considering Second Order Derivative

Sheng Lei and Alexander Flueck
Department of Electrical and Computer Engineering, Illinois Institute of Technology, Chicago, IL, USA
Email: slei3@hawk.iit.edu and flueck@iit.edu

*Abstract*—Potential disagreement in the result induced by discontinuities is revealed in this paper between a novel power system transient simulation scheme using numerical integrators considering second order derivative and conventional ones using numerical integrators considering first order derivative. The disagreement is due to the formula of the different numerical integrators. An improved method for dealing with discontinuities in the novel transient simulation scheme is proposed to resolve the disagreement. The effectiveness of the improved method is demonstrated and verified via numerical case studies. Although the disagreement is studied on and the improved method is proposed for a particular transient simulation scheme, similar conclusions also apply to other ones using numerical integrators considering high order derivative.

## I. INTRODUCTION

Discontinuities frequently happen in electronic circuits and power systems, typically in the form of switching operations and short-circuit faults. They should be properly dealt with to avoid incorrect results in transient simulations. Conventionally, transient simulation schemes use numerical integrators considering first order derivative, such as the implicit trapezoidal method (ITM) and the 2-step backward differentiation formula (BDF2), to discretize the differential equations of the studied circuits and systems at normal time steps [1]-[3]. When a discontinuity occurs, the numerical integrator is temporarily switched to the backward Euler method (BEM) for several time steps [1], [3].

To simultaneously achieve accuracy and efficiency in transient simulation, some schemes using numerical integrators considering high order derivative are proposed in the literature [4]-[5]. In [4], Obreshkov numerical integrators are introduced into circuit simulation. In [5], frequency response optimized integrators considering second order derivative are defined; and a novel power system transient simulation scheme is put forward based on the proposed numerical integrators. However, to the authors' knowledge, how to deal with discontinuities in transient simulation based on numerical integrators considering high order derivative is not studied in detail in the literature. In [5], a preliminary method is adopted, where numerical integrators are used for two half-steps immediately after a discontinuity that do not rely on information at the previous time steps except for differential state variables. The preliminary method is motivated by the critical damping adjustment (CDA) [3] commonly used in power system transient simulation, where two half-steps of BEM are temporarily used in place of ITM immediately after a discontinuity.

Further studies show that the novel transient simulation scheme with the preliminary method for dealing with discontinuities [5] may disagree with conventional transient simulation schemes using numerical integrators considering first order derivative in some cases, although they usually reach an agreement. The disagreement will be reported in this paper. It is desirable for the results from the novel transient simulation scheme to match those from conventional schemes, because novel schemes are usually compared and validated with conventional ones [4]-[5]. If the results do not agree, the validity of a novel scheme may seem doubtful to a user.

This paper studies the potential disagreement in the result between the novel transient simulation scheme [5] and conventional ones as well as a mitigation strategy. The main contributions of the paper are twofold. First, the cause of the disagreement is identified, which is due to the formula of the different numerical integrators. Second, an improved method for dealing with discontinuities in the novel transient simulation scheme is proposed to resolve the disagreement. The effectiveness of the proposed improved method is demonstrated and verified via numerical case studies. The rest of the paper is organized as follows. Section II introduces necessary background information regarding numerical solution of a differential-algebraic equation (DAE) set, which is the typical mathematical model of electronic circuits and power systems [1]-[2], [6]-[7]. Section III studies in detail calculations immediately after a discontinuity in the novel transient simulation scheme [5] and conventional ones with a concrete example, reveals the disagreement between the respective results and identifies the cause. Section IV proposes the improved method for dealing with discontinuities. Section V demonstrates and verifies the effectiveness of the improved method via numerical case studies. Section VI concludes the paper.

## II. NUMERICAL SOLUTION TO DAE SET

Consider a general DAE set

$$\begin{cases} \dot{x} = f(x, y, \tau) \\ 0 = g(x, y, \tau) \end{cases} \quad (1)$$

where $x$ is the differential state variable; $y$ is the algebraic state variable; $\tau$ is the time variable; $f$ and $g$ are functions depending on $x$, $y$ and $\tau$. Numerical integrators considering first order derivative may be applied to solve (1)

$$\begin{aligned} x_{t+h} &= x_t + b_0 \dot{x}_{t+h} + b_{-1} \dot{x}_t \\ 0 &= g(x, y, \tau)|_{t+h} \\ (\dot{x} &= f(x, y, \tau)) \end{aligned} \quad (2)$$

where $t$ is the current time instant; $h$ is the step size; $b_0$ and $b_{-1}$ are coefficients of the numerical integrator. For example, BEM has $b_0 = h$, $b_{-1} = 0$ while ITM has $b_0 = h/2$, $b_{-1} = h/2$. It is assumed that the time step at $t + h$ is to be solved while the time step at $t$ is known. Numerical integrators considering second order derivative may also be applied to solve (1)

$$\begin{aligned} x_{t+h} &= x_t + b_0 \dot{x}_{t+h} + b_{-1} \dot{x}_t + c_0 \ddot{x}_{t+h} + c_{-1} \ddot{x}_t \\ 0 &= g(x, y, \tau)|_{t+h} \\ 0 &= (\frac{\partial g}{\partial x} \dot{x} + \frac{\partial g}{\partial y} \dot{y} + \frac{\partial g}{\partial \tau})|_{t+h} \\ (\dot{x} &= f(x, y, \tau),\ \ddot{x} = \frac{\partial f}{\partial x} \dot{x} + \frac{\partial f}{\partial y} \dot{y} + \frac{\partial f}{\partial \tau}) \end{aligned} \quad (3)$$

where $b_0$, $b_{-1}$, $c_0$ and $c_{-1}$ are coefficients of the numerical integrator. Several numerical integrators of this type are given in [5]. Note that $\dot{y}$ is additionally introduced as an unknown to be solved in (3). Similarly, numerical integrators considering high order derivative [8] are applicable.

## III. DISAGREEMENT BETWEEN TRANSIENT SIMULATION SCHEMES

This section takes a close look at calculations following a discontinuity in a transient simulation run. Disagreement in the result between the novel transient simulation scheme [5] and conventional ones using numerical integrators considering first order derivative [1]-[2] is revealed.

Consider a simple circuit as shown in Fig. 1. Parameters and variables of the circuit are marked in the figure. Suppose that the voltage at the middle point $u$ is nonzero at a time instant $t$, namely $u_t \neq 0$. Accordingly, the current going through the resistor is nonzero. According to Kirchhoff's Current Law (KCL)

$$i_{1,t} + i_{2,t} \neq 0 \quad (4)$$

Further suppose that the time step at $t$ has just been solved; and the switch is opened at $t$. The time step immediately after this discontinuity is to be calculated. The mathematical model of the circuit after the discontinuity is the DAE set

$$L_1 \frac{di_1}{dt} = u - v_1,\ L_2 \frac{di_2}{dt} = u - v_2,\ i_1 + i_2 = 0 \quad (5)$$

When using numerical integrators considering second order derivative to solve (5), $\dot{v}_1$ and $\dot{v}_2$ are supposed to be given.

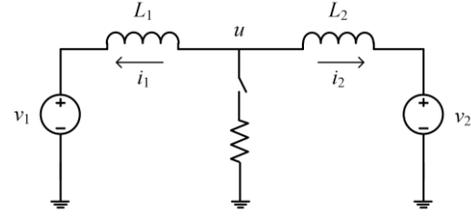

Fig. 1. Simple circuit.

### A. Using BEM

Conventional transient simulation schemes use BEM for discontinuities [1], [3]. The DAE set (5) is discretized by BEM as

$$\begin{aligned} i_{1,t+h} &= i_{1,t} + b_0 \frac{1}{L_1}(u_{t+h} - v_{1,t+h}) \\ i_{2,t+h} &= i_{2,t} + b_0 \frac{1}{L_2}(u_{t+h} - v_{2,t+h}) \\ i_{1,t+h} &+ i_{2,t+h} = 0 \end{aligned} \quad (6)$$

Note that $b_{-1} = 0$ for BEM so the corresponding terms are skipped in (6). Comparing (2) and (6), it is learnt that BEM does not rely on information at the previous time step except for differential state variables, which do not suffer from instantaneous change in the value [6]-[7]. Therefore the method is suitable for dealing with discontinuities that may induce instantaneous change of some variables. According to (6), $u$ at $t + h$ is calculated as

$$u_{t+h} = \frac{L_2 v_{1,t+h} + L_1 v_{2,t+h}}{L_1 + L_2} - \frac{L_1 L_2}{L_1 + L_2} \frac{1}{b_0}(i_{1,t} + i_{2,t}) \quad (7)$$

It is observed in (7) that there is a step size dependent excursion in $u$ which is

$$-\frac{L_1 L_2}{L_1 + L_2} \frac{1}{b_0}(i_{1,t} + i_{2,t}) \quad (8)$$

### B. Using Numerical Integrators Considering Second Order Derivative with Zero $b_{-1}$ and $c_{-1}$

In the novel transient simulation scheme, numerical integrators with zero $b_{-1}$ and $c_{-1}$ are used immediately after discontinuities [5]. Similar to BEM, these numerical integrators skip the potential instantaneous change in some variables and are thus suitable for discontinuities. When applying one such numerical integrator, (5) is discretized as

$$\begin{aligned} i_{1,t+h} &= i_{1,t} + b_0 \frac{1}{L_1}(u_{t+h} - v_{1,t+h}) + c_0 \frac{1}{L_1}(\dot{u}_{t+h} - \dot{v}_{1,t+h}) \\ i_{2,t+h} &= i_{2,t} + b_0 \frac{1}{L_2}(u_{t+h} - v_{2,t+h}) + c_0 \frac{1}{L_2}(\dot{u}_{t+h} - \dot{v}_{2,t+h}) \\ i_{1,t+h} &+ i_{2,t+h} = 0 \\ \dot{i}_{1,t+h} &+ \dot{i}_{2,t+h} = 0 \\ (L_1 \dot{i}_1 &= u - v_1,\ L_2 \dot{i}_2 = u - v_2) \end{aligned} \quad (9)$$

The terms corresponding to the zero $b_{-1}$ and $c_{-1}$ are skipped in (9). $u$ and $\dot{u}$ are calculated as

$$\begin{aligned} u_{t+h} &= \frac{L_2 v_{1,t+h} + L_1 v_{2,t+h}}{L_1 + L_2} \\ \dot{u}_{t+h} &= \frac{L_2 \dot{v}_{1,t+h} + L_1 \dot{v}_{2,t+h}}{L_1 + L_2} - \frac{L_1 L_2}{L_1 + L_2} \frac{1}{c_0}(i_{1,t} + i_{2,t}) \end{aligned} \quad (10)$$

Note in (10) that there is no step size dependent excursion in $u$ but one in $\dot{u}$ which is

$$-\frac{L_1 L_2}{L_1 + L_2}\frac{1}{c_0}(i_{1,t} + i_{2,t}) \qquad (11)$$

*C. Comments*

Comparing (7) and (10), it is observed that there is a difference in the calculated voltage from the different transient simulation schemes. This difference is not caused by mistakes in the implementation, but attributed to the formula of the different numerical integrators. If the voltage is used as an input signal to certain controllers, such as a voltage regulator, the difference will trigger cascading reactions of the studied system, leading to significant disagreement between the overall results from the different transient simulation schemes. A demonstrative example about the disagreement will be presented in Section V.

Similar conclusion holds for other transient simulation schemes using numerical integrators considering high order derivative. There will be a step size dependent excursion in a high order derivative of the voltage but there will be none in the voltage itself and its lower order derivatives. The overall results may disagree with those from conventional transient simulation schemes.

## IV. PROPOSED IMPROVED METHOD

*A. Description of the Method*

It is desirable for the results from the novel transient simulation scheme [5] to match the results from conventional ones for comparison and validation. Considering the cause of the disagreement detailed in the previous section, the first time step immediately after a discontinuity should be calculated with BEM, following the conventional approach. To reduce the error introduced by BEM, the step size should be shrunk to a small number. This is achieved by setting $b_0 = \varepsilon$ and other coefficients to be zero in (3), where $\varepsilon$ is a user-defined small number. As $c_0$ is set to zero, $\dot{y}$ may become unsolvable. This is exactly the case in (9), where $\dot{u}_{t+h}$ is unsolvable with zero $c_0$. To resolve the insolvability while avoiding eliminating some variables which will complicate the solution process, $\dot{y}$ is temporarily set to zero at this time step. Note that this artificial setting does not impact the solution to other variables because BEM does not consider such information.

The next time step is to be calculated with numerical integrators considering second order derivative with zero $b_{-1}$ and $c_{-1}$ [5] to reinitialize $\dot{y}$. As $c_0$ is nonzero at this time step, $\dot{y}$ is solvable. The zero $b_{-1}$ and $c_{-1}$ skip the artificial value of $\dot{y}$ at the previous time step so that it does not impact the calculation at the current time step. Note that numerical integrators with zero $b_{-1}$ and $c_{-1}$ are less accurate [5]. The step size of the second time step immediately after a discontinuity should also be reduced to ensure accuracy.

This paper proposes an arrangement which calculates the second and third time steps immediately after a discontinuity using numerical integrators considering second order derivative with zero $b_{-1}$ and $c_{-1}$, with a reduced step size of $(h - \varepsilon) / 2$. After the aforementioned three small time steps, the simulation completes a full time step and can be carried on with the normal step size $h$ and normal numerical integrators.

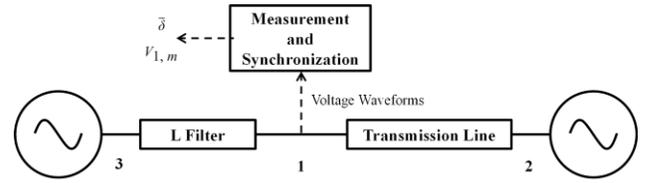

Fig. 2. Single-line diagram of the test system.

*B. Summary*

The proposed method for dealing with discontinuities in the novel transient simulation scheme consists of the following steps:
1) At the first time step immediately after a discontinuity, (3) is solved by setting $b_0 = \varepsilon$ and $b_{-1} = c_0 = c_{-1} = 0$; the step size is $\varepsilon$; $\dot{y}$ is artificially set to zero.
2) At the second and third time steps immediately after a discontinuity, (3) is solved with numerical integrators considering second order derivative which have zero $b_{-1}$ and $c_{-1}$; the step size is $(h - \varepsilon) / 2$.
3) The intermediate time steps, namely the first and second time steps immediately after a discontinuity, are calculated but the results are not output to avoid confusion; once a full time step consisting of the three small time steps is completed, the simulation is carried on with normal numerical integrators and a step size of $h$.

Although the above method is proposed for the novel transient simulation scheme, it can be extended to other ones using numerical integrators considering high order derivative in a straightforward manner.

## V. NUMERICAL CASE STUDIES

*A. Description of the Test System*

The novel transient simulation scheme is applied to a test system in this section with its previous method for dealing with discontinuities [5] and the improved method proposed in this paper. The simulation results from the novel transient simulation scheme are compared to those from an iterative electromagnetic transient simulator (iEMTS) [5] applied to the same test system. The iEMTS is conventional in that it uses ITM at normal time steps and two half-steps of BEM for discontinuities.

The test system is a three-bus three-phase power system. Its single-line diagram is shown in Fig. 2. Following the convention in power system engineering, parameters and variables are expressed in per-unit (p.u.) [9] in this paper, except for angles and time. A three-phase transmission line connects Busses 1 and 2. The transmission line is modeled as a decoupled three-phase impedance; the per-phase impedance is $0.0529 + j0.4288$. An L filter connects Busses 3 and 1, which is modeled as a decoupled three-phase inductor. The per-phase inductance is 0.16. A three-phase AC voltage source is connected to Bus 3; its voltage magnitude and angle are 1.0131 and 0.5834 rad respectively. Another three-phase AC voltage source is connected to Bus 2; its voltage magnitude and angle are 1.04 and 0.0 rad respectively. Both voltage sources are ideal; if the specified voltage magnitude and angle are $V_m$ and $\theta$ respectively, the Phase A, B and C voltage outputs are

$$V_m \cos(120\pi t + \theta),\ V_m \cos(120\pi t + \theta - \frac{2}{3}\pi),\ V_m \cos(120\pi t + \theta + \frac{2}{3}\pi) \quad (12)$$

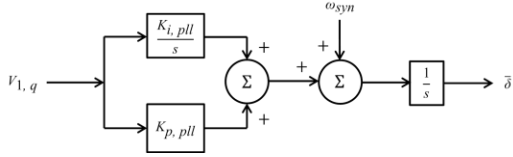

Fig. 3. Phase-locked loop (PLL).

A measurement and synchronization mechanism measures some electrical quantities at Bus 1, which will be detailed in the coming subsection.

Simulation runs start at 0.0 s from the steady state. At 0.2 s, a three-phase-to-ground fault is applied at Bus 1 with a fault resistance of 0.1. At 0.4 s, the fault is cleared.

### B. Measurement and Synchronization Mechanism

*1) In-phase and quadrature signal generation:* Applying Clarke's transformation to the three-phase components of Bus 1 voltage, the in-phase and quadrature signals are calculated as

$$\begin{pmatrix} v_{1,in} \\ v_{1,qu} \end{pmatrix} = \frac{2}{3} \begin{pmatrix} \cos(0) & \cos(-\frac{2}{3}\pi) & \cos(\frac{2}{3}\pi) \\ -\sin(0) & -\sin(-\frac{2}{3}\pi) & -\sin(\frac{2}{3}\pi) \end{pmatrix} \begin{pmatrix} v_{1,A} \\ v_{1,B} \\ v_{1,C} \end{pmatrix} \quad (13)$$

where $v_{1,A}$, $v_{1,B}$ and $v_{1,C}$ are the Phase A, B and C component of Bus 1 voltage respectively; $v_{1,in}$ and $v_{1,qu}$ are the in-phase and quadrature signal of Bus 1 voltage respectively.

*2) Phase shift:* A phase shift factor is applied to $v_{1,in}$ and $v_{1,qu}$ to calculate the real and imaginary parts of Bus 1 voltage phasor in the device reference frame

$$V_1 = V_{1,d} + jV_{1,q} = e^{-j\bar{\delta}}(v_{1,in} + jv_{1,qu}) \quad (14)$$

where $V_1$ is Bus 1 voltage phasor; $V_{1,d}$ and $V_{1,q}$ are the real and imaginary part of Bus 1 voltage phasor respectively; $\bar{\delta}$ is the cumulative phase angle of Bus 1 voltage.

*3) Phase-locked loop (PLL):* A PLL is shown in Fig. 3. $V_{1,q}$ is sent to the PLL as the input. The output is $\bar{\delta}$. $\omega_{syn}$ is the synchronous angular frequency.

*4) Bus 1 voltage phasor magnitude measurement:* Bus 1 voltage phasor magnitude is first pre-calculated as

$$V_{1,m,pre} = \sqrt{V_{1,d}^2 + V_{1,q}^2} \quad (15)$$

where $V_{1,m,pre}$ is the pre-calculated Bus 1 voltage phasor magnitude. The measured value is then obtained by passing the pre-calculated value to a low-pass filter

$$\dot{V}_{1,m} = \frac{1}{T_V}(-V_{1,m} + V_{1,m,pre}) \quad (16)$$

where $V_{1,m}$ is the measured Bus 1 voltage phasor magnitude; $T_V$ is the time constant of the low-pass filter.

### C. Results

The measured Bus 1 voltage phasor magnitude from the iEMTS with a tiny step size of 5 μs is used as the reference.

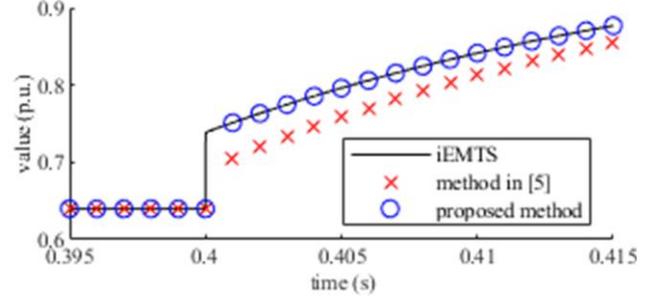

Fig. 4. Measured Bus 1 voltage phasor magnitude.

Results from the novel transient simulation scheme using different methods for dealing with discontinuities are compared to the reference adopting a step size of 1 ms in Fig. 4. For better visualization, only the comparison around 0.405 s is presented. Results from the novel transient simulation scheme closely match the reference before the discontinuity at 0.4 s. Nevertheless, the method in [5] disagrees with the reference after the discontinuity, as discussed in Section III.C. The proposed improved method reaches a much closer agreement with the reference than the method in [5]. The effectiveness of the proposed improved method is thus verified.

## VI. CONCLUSION

Potential disagreement induced by discontinuities between the novel transient simulation scheme [5] and conventional ones is revealed, which is due to the formula of the different numerical integrators. An improved method for dealing with discontinuities is proposed and validated to resolve the disagreement. Similar improvements can be readily extended to other transient simulation schemes using numerical integrators considering high order derivative.